\pdfoutput=1
\documentclass[12pt,a4paper]{article}
\usepackage{ifthen} % for conditional statements
\newboolean{pdflatex}
\setboolean{pdflatex}{true} % False for eps figures 

\newboolean{articletitles}
\setboolean{articletitles}{true} % False removes titles in references

\newboolean{uprightparticles}
\setboolean{uprightparticles}{false} %True for upright particle symbols

\def\paperauthors{LHCb collaboration} % Leave as is for PAPER, CONF and FIGURE
\def\paperasciititle{Searches for B0->Kpitautau and Bs->KKtautau decays}
\def\papertitle{Searches for $\Bz\to\Kp\pim\taup\taum$ and $\Bs\to\Kp\Km\taup\taum$ decays}
\def\paperkeywords{{High Energy Physics}, {LHCb}} % Comma separated list
\def\papercopyright{\the\year\ CERN for the benefit of the LHCb collaboration}
\def\paperlicence{CC BY 4.0 licence}
\def\paperlicenceurl{https://creativecommons.org/licenses/by/4.0/}

\newif\ifEnableSectionTOCLinks
\EnableSectionTOCLinksfalse % deactivated

\usepackage[top=1in, bottom=1.25in, left=1in, right=1in]{geometry}

\usepackage{microtype}
\usepackage{lineno}  % for line numbering during review
\usepackage{xspace} % To avoid problems with missing or double spaces after
\usepackage{caption} %these three command get the figure and table captions automatically small

\usepackage{graphicx}  % to include figures (can also use other packages)
\usepackage{color}
\usepackage{colortbl}
\graphicspath{{./figs/}} % Make Latex search fig subdir for figures
\usepackage{amsmath} % Adds a large collection of math symbols
\usepackage{amssymb}
\usepackage{amsfonts}
\usepackage{upgreek} % Adds in support for greek letters in roman typeset

\newcommand*\patchAmsMathEnvironmentForLineno[1]{%
\expandafter\let\csname old#1\expandafter\endcsname\csname #1\endcsname
\expandafter\let\csname oldend#1\expandafter\endcsname\csname
end#1\endcsname
 \renewenvironment{#1}%
   {\linenomath\csname old#1\endcsname}%
   {\csname oldend#1\endcsname\endlinenomath}%
}
\newcommand*\patchBothAmsMathEnvironmentsForLineno[1]{%
  \patchAmsMathEnvironmentForLineno{#1}%
  \patchAmsMathEnvironmentForLineno{#1*}%
}
\AtBeginDocument{%
\patchBothAmsMathEnvironmentsForLineno{equation}%
\patchBothAmsMathEnvironmentsForLineno{align}%
\patchBothAmsMathEnvironmentsForLineno{flalign}%
\patchBothAmsMathEnvironmentsForLineno{alignat}%
\patchBothAmsMathEnvironmentsForLineno{gather}%
\patchBothAmsMathEnvironmentsForLineno{multline}%
\patchBothAmsMathEnvironmentsForLineno{eqnarray}%
}

\usepackage[pdftex,
            pdfauthor={\paperauthors},
            pdftitle={\paperasciititle},
            pdfkeywords={\paperkeywords}]{hyperref}
\usepackage{hyperxmp}
\hypersetup{
    pdfcopyright={Copyright (C) \papercopyright},
    pdflicenseurl={\paperlicenceurl}
}
\usepackage[colorinlistoftodos,textsize=scriptsize]{todonotes}

\usepackage[bottom,flushmargin,hang,multiple]{footmisc}

\usepackage[all]{hypcap} % Internal hyperlinks to floats.

\usepackage{xspace} 
\usepackage{upgreek}

\def\lhcb   {\mbox{LHCb}\xspace}

\def\MagUp {\mbox{\em Mag\kern -0.05em Up}\xspace}

\ifthenelse{\boolean{uprightparticles}}%
{

 \def\Pmu         {\ensuremath{\upmu}\xspace}                 
 \def\Pnu         {\ensuremath{\upnu}\xspace}                 
                  
 \def\Ppi         {\ensuremath{\uppi}\xspace}

 \def\Ptau        {\ensuremath{\uptau}\xspace}                 
                  
 \def\Pphi        {\ensuremath{\upphi}\xspace}

 \def\Ppsi        {\ensuremath{\uppsi}\xspace}

 \def\PDelta      {\ensuremath{\Delta}\xspace}                 
 \def\PXi         {\ensuremath{\Xi}\xspace}                 
 \def\PLambda     {\ensuremath{\Lambda}\xspace}                 
 \def\PSigma      {\ensuremath{\Sigma}\xspace}                 
 \def\POmega      {\ensuremath{\Omega}\xspace}                 
 \def\PUpsilon    {\ensuremath{\Upsilon}\xspace}
 \let\oldPi\Pi
 \def\PPi         {\ensuremath{\oldPi}\xspace}

 \def\PB      {\ensuremath{\mathrm{B}}\xspace}                 
 \def\PD      {\ensuremath{\mathrm{D}}\xspace}                 
 \def\PJ      {\ensuremath{\mathrm{J}}\xspace}                 
 \def\PK      {\ensuremath{\mathrm{K}}\xspace}                 
 \def\Pb      {\ensuremath{\mathrm{b}}\xspace}                 
 \def\Pc      {\ensuremath{\mathrm{c}}\xspace}

 \def\Ph      {\ensuremath{\mathrm{h}}\xspace}                 
 \def\Ps      {\ensuremath{\mathrm{s}}\xspace}

 \def\thebaroffset{0.0em}
}
{

 \def\Pmu         {\ensuremath{\mu}\xspace}                 
 \def\Pnu         {\ensuremath{\nu}\xspace}                 
                  
 \def\Ppi         {\ensuremath{\pi}\xspace}

 \def\Ptau        {\ensuremath{\tau}\xspace}                 
                  
 \def\Pphi        {\ensuremath{\phi}\xspace}

 \def\Ppsi        {\ensuremath{\psi}\xspace}                 
                  
 \mathchardef\PDelta="7101
 \mathchardef\PXi="7104
 \mathchardef\PLambda="7103
 \mathchardef\PSigma="7106
 \mathchardef\POmega="710A
 \mathchardef\PUpsilon="7107
 \mathchardef\PPi="7105
 \def\PB      {\ensuremath{B}\xspace}                 
 \def\PD      {\ensuremath{D}\xspace}                 
 \def\PJ      {\ensuremath{J}\xspace}                 
 \def\PK      {\ensuremath{K}\xspace}                 
 \def\Pb      {\ensuremath{b}\xspace}                 
 \def\Pc      {\ensuremath{c}\xspace}

 \def\Ph      {\ensuremath{h}\xspace}                 
 \def\Ps      {\ensuremath{s}\xspace}

 \def\thebaroffset{0.18em}
}
\newcommand{\offsetoverline}[2][\thebaroffset]{\kern #1\overline{\kern -#1 #2}}%

\makeatletter
\ifcase \@ptsize \relax% 10pt
  \newcommand{\miniscule}{\@setfontsize\miniscule{4}{5}}% \tiny: 5/6
\or% 11pt
  \newcommand{\miniscule}{\@setfontsize\miniscule{5}{6}}% \tiny: 6/7
\or% 12pt
  \newcommand{\miniscule}{\@setfontsize\miniscule{5}{6}}% \tiny: 6/7
\fi
\makeatother

\DeclareRobustCommand{\optbar}[1]{\shortstack{{\miniscule (\rule[.5ex]{1.25em}{.18mm})}
  \\ [-.7ex] $#1$}}

\def\mup        {{\ensuremath{\Pmu^+}}\xspace}
\def\mun        {{\ensuremath{\Pmu^-}}\xspace} % muon negative (\mum is taken)

\def\taup       {{\ensuremath{\Ptau^+}}\xspace}
\def\taum       {{\ensuremath{\Ptau^-}}\xspace}

\def\neu        {{\ensuremath{\Pnu}}\xspace}
\def\neub       {{\ensuremath{\overline{\Pnu}}}\xspace}
\def\neum       {{\ensuremath{\neu_\mu}}\xspace}
\def\neumb      {{\ensuremath{\neub_\mu}}\xspace}

\def\neutb      {{\ensuremath{\neub_\tau}}\xspace}
\def\squark    {{\ensuremath{\Ps}}\xspace}

\def\cquark    {{\ensuremath{\Pc}}\xspace}

\def\bquark    {{\ensuremath{\Pb}}\xspace}

\def\hadron {{\ensuremath{\Ph}}\xspace}
\def\pion   {{\ensuremath{\Ppi}}\xspace}

\def\pip    {{\ensuremath{\pion^+}}\xspace}
\def\pim    {{\ensuremath{\pion^-}}\xspace}

\def\kaon    {{\ensuremath{\PK}}\xspace}
\def\KorKbar {\kern \thebaroffset\optbar{\kern -\thebaroffset \PK}{}\xspace}
\def\Kz      {{\ensuremath{\kaon^0}}\xspace}

\def\Kp      {{\ensuremath{\kaon^+}}\xspace}
\def\Km      {{\ensuremath{\kaon^-}}\xspace}

\def\Kstarz  {{\ensuremath{\kaon^{*0}}}\xspace}

\def\Kstar   {{\ensuremath{\kaon^*}}\xspace}

\newcommand{\phiz}{\ensuremath{\Pphi}\xspace}

\def\Dbar    {{\ensuremath{\offsetoverline{\PD}}}\xspace}
\def\D       {{\ensuremath{\PD}}\xspace}
\def\Db      {{\ensuremath{\Dbar}}\xspace}
\def\DorDbar {\kern \thebaroffset\optbar{\kern -\thebaroffset \PD}\xspace}

\def\Dzb     {{\ensuremath{\Dbar{}^0}}\xspace}
\def\Dp      {{\ensuremath{\D^+}}\xspace}
\def\Dm      {{\ensuremath{\D^-}}\xspace}

\def\DpDm    {\ensuremath{\Dp {\kern -0.16em \Dm}}\xspace}

\def\Dsm     {{\ensuremath{\D^-_\squark}}\xspace}

\def\B       {{\ensuremath{\PB}}\xspace}

\def\BorBbar {\kern \thebaroffset\optbar{\kern -\thebaroffset \PB}\xspace}
\def\Bz      {{\ensuremath{\B^0}}\xspace}

\def\Bd      {{\ensuremath{\B^0}}\xspace}

\def\BdorBdbar {\kern \thebaroffset\optbar{\kern -\thebaroffset \Bd}\xspace}
\def\Bu      {{\ensuremath{\B^+}}\xspace}

\def\Bp      {{\ensuremath{\Bu}}\xspace}

\def\Bs      {{\ensuremath{\B^0_\squark}}\xspace}

\def\BsorBsbar {\kern \thebaroffset\optbar{\kern -\thebaroffset \Bs}\xspace}

\def\Bds     {{\ensuremath{\B_{(\squark)}^0}}\xspace}

\def\BdorBs  {\Bds}

\def\jpsi     {{\ensuremath{{\PJ\mskip -3mu/\mskip -2mu\Ppsi}}}\xspace}

\def\Y#1S{\ensuremath{\PUpsilon{(#1S)}}\xspace}

\def\LorLbar     {\kern \thebaroffset\optbar{\kern -\thebaroffset \PLambda}\xspace}

\newcommand{\decay}[2]{\mbox{\ensuremath{#1\!\to #2}}\xspace} 

\def\to                 {\ensuremath{\rightarrow}\xspace}

\def\qsq       {{\ensuremath{q^2}}\xspace}

\def\AT#1     {\ensuremath{A_{\mathrm{T}}^{#1}}\xspace}           % 2

\def\C#1      {\ensuremath{\mathcal{C}_{#1}}\xspace}                       % 9
\def\Cp#1     {\ensuremath{\mathcal{C}_{#1}^{'}}\xspace}                    % 7
\def\Ceff#1   {\ensuremath{\mathcal{C}_{#1}^{\mathrm{(eff)}}}\xspace}        % 9  
\def\Cpeff#1  {\ensuremath{\mathcal{C}_{#1}^{'\mathrm{(eff)}}}\xspace}       % 7
\def\Ope#1    {\ensuremath{\mathcal{O}_{#1}}\xspace}                       % 2
\def\Opep#1   {\ensuremath{\mathcal{O}_{#1}^{'}}\xspace}                    % 7

\newcommand{\aunit}[1]{\ensuremath{\text{\,#1}}}       
\newcommand{\tev}{\aunit{Te\kern -0.1em V}\xspace}
\newcommand{\gev}{\aunit{Ge\kern -0.1em V}\xspace}
\newcommand{\mev}{\aunit{Me\kern -0.1em V}\xspace}
\newcommand{\kev}{\aunit{ke\kern -0.1em V}\xspace}
\newcommand{\ev}{\aunit{e\kern -0.1em V}\xspace}
 
\newcommand{\mevc}{\ensuremath{\aunit{Me\kern -0.1em V\!/}c}\xspace}
\newcommand{\gevc}{\ensuremath{\aunit{Ge\kern -0.1em V\!/}c}\xspace}
\newcommand{\mevcc}{\ensuremath{\aunit{Me\kern -0.1em V\!/}c^2}\xspace}
\newcommand{\gevcc}{\ensuremath{\aunit{Ge\kern -0.1em V\!/}c^2}\xspace}
\def\fb   {\ensuremath{\aunit{fb}}\xspace}
\def\invfb   {\ensuremath{\fb^{-1}}\xspace}

\def\gsim{{~\raise.15em\hbox{$>$}\kern-.85em
          \lower.35em\hbox{$\sim$}~}\xspace}
\def\lsim{{~\raise.15em\hbox{$<$}\kern-.85em
          \lower.35em\hbox{$\sim$}~}\xspace}

\def\evtgen     {\mbox{\textsc{EvtGen}}\xspace}

\def\geant      {\mbox{\textsc{Geant4}}\xspace}

\def\photos     {\mbox{\textsc{Photos}}\xspace}

\def\pythia     {\mbox{\textsc{Pythia}}\xspace}

\def\tell1  {TELL1\xspace}
\def\ukl1   {UKL1\xspace}

\newcommand{\lhcborcid}[1]{\href{https://orcid.org/#1}{\hspace*{0.1em}\raisebox{-0.45ex}{\includegraphics[width=1em]{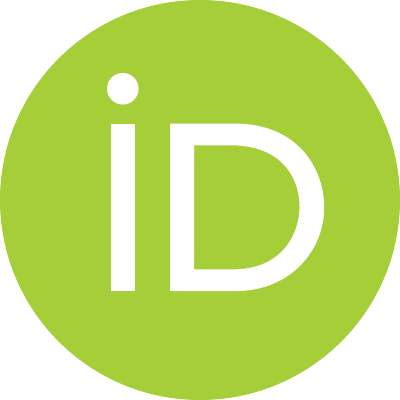}}}}

\hypersetup{
  colorlinks   = true, %Colours links instead of ugly boxes
  urlcolor     = blue, %Colour for external hyperlinks
  linkcolor    = blue, %Colour of internal links
  citecolor    = red   %Colour of citations
}

\ifEnableSectionTOCLinks
    \usepackage[explicit]{titlesec} % to change headings
    
    \let\oldcontentsline\contentsline
    \renewcommand

    \titleformat{\section}{\normalfont\Large\bf}{\hyperlink{tocsection.\thesection}{{\thesection} \parbox[t]{\dimexpr\textwidth-1pc}{#1}}}{1pc}{}

    \titleformat{\subsection}{\normalfont\bf}{\hyperlink{tocsubsection.\thesubsection}{{\thesubsection} \parbox[t]{\dimexpr\textwidth-1pc}{#1}}}{1pc}{}

    \titleformat{name=\section,numberless}[display]{}{}{0pt}{\normalfont\Huge\bfseries #1}
\fi

\usepackage{cite} % Allows for ranges in citations
\usepackage{mciteplus}

\usepackage{booktabs}
\usepackage{placeins}
\usepackage{pdflscape}

\usepackage{verbatim}

\begin{document}

%%%%%%%%%%%%%%%%%%%%%%%%%
%%%%% Title     %%%%%%%%%
%%%%%%%%%%%%%%%%%%%%%%%%%
\renewcommand{\thefootnote}{\fnsymbol{footnote}}
\setcounter{footnote}{1}

% %%%%%%% CHOOSE TITLE PAGE--------
%\onecolumn

% ===============================================================================
% Purpose: LHCb-PAPER journal paper title page template
% Author: 
% Created on: 2010-09-25
% ===============================================================================

%%%%%%%%%%%%%%%%%%%%%%%%%
%%%%%  TITLE PAGE  %%%%%%
%%%%%%%%%%%%%%%%%%%%%%%%%
\begin{titlepage}
\pagenumbering{roman}

% Header ---------------------------------------------------
\vspace*{-1.5cm}
\centerline{\large EUROPEAN ORGANIZATION FOR NUCLEAR RESEARCH (CERN)}
\vspace*{1.5cm}
\noindent
\begin{tabular*}{\linewidth}{lc@{\extracolsep{\fill}}r@{\extracolsep{0pt}}}
\ifthenelse{\boolean{pdflatex}}% Logo format choice
{\vspace*{-1.5cm}\mbox{\!\!\!\includegraphics[width=.14\textwidth]{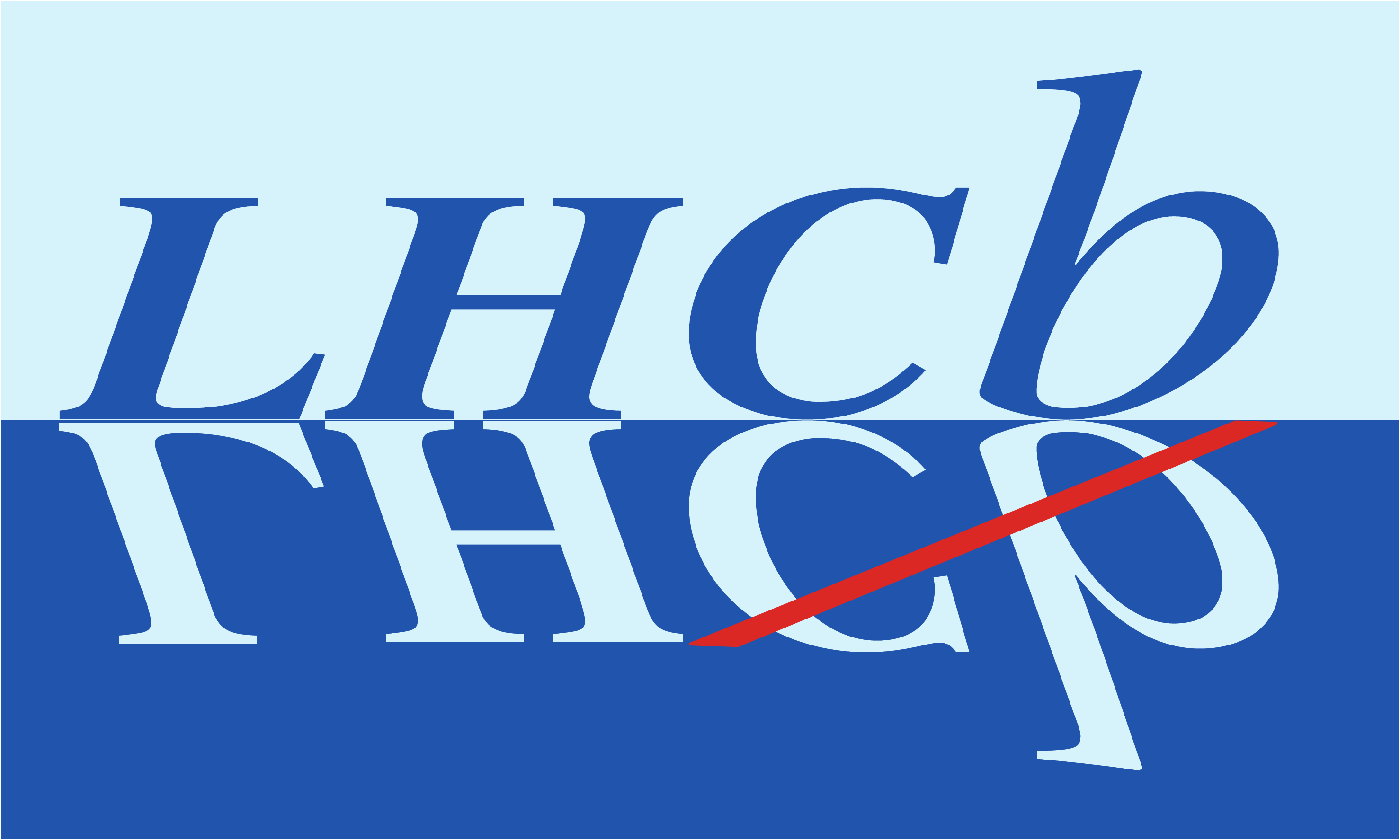}} & &}%
{\vspace*{-1.2cm}\mbox{\!\!\!\includegraphics[width=.12\textwidth]{LHCb/lhcb-logo.eps}} & &}%
\\
 & & CERN-EP-2025-224 \\  
 & & LHCb-PAPER-2025-048 \\ 
 & & May 5, 2026 \\ 
 & & \\
\end{tabular*}

\vspace*{4.0cm}

% Title --------------------------------------------------
{\normalfont\bfseries\boldmath\huge
\begin{center}
  \papertitle 
\end{center}
}

% Authors -------------------------------------------------
\begin{center}
\paperauthors\footnote{Authors are listed at the end of this Letter.}
\end{center}

\vspace{\fill}

% Abstract -----------------------------------------------
\begin{abstract}
  \noindent
  The first searches for $B^0\to K^+\pi^-\tau^+\tau^-$ and $B^0_s\to K^+K^-\tau^+\tau^-$  decays at the LHCb experiment are conducted with $pp$ collision data corresponding to an integrated luminosity of $5.4\invfb$. The tau leptons are reconstructed using the \mbox{$\tau^+\to \mu^+\overline{\nu}_\tau\nu_\mu$} decay and the results are presented in bins of $K^+\pi^-$ or $K^+K^-$ mass. No signal is observed and upper limits are set on the branching fractions. The searches result in the first upper limits for $B^0\to K^+\pi^-\tau^+\tau^-$ decays outside the $K^*(892)^0$  region  in $K^+\pi^-$ mass and the first limits for $B^0_s\to K^+K^-\tau^+\tau^-$ decays. The searches are recast into limits on the decays $B^0\to K^*(892)^0\tau^+\tau^-$ and \mbox{$B^0_s\to \phi(1020)\tau^+\tau^-$}, yielding \mbox{$2.8\times10^{-4}$ ($2.5\times10^{-4}$)} and \mbox{$4.7\times10^{-4}$ ($4.1\times10^{-4}$)} at the \mbox{$95\%$ ($90\%$)} confidence level, respectively. For the decay \mbox{$B^0\to K^*(892)^0\tau^+\tau^-$}, this result improves on the current best upper limit by an order of magnitude.   
\end{abstract}

\vspace*{2.0cm}

\begin{center}
  Published in Physical Review Letters 136 (2026) 181802 
\end{center}

\vspace{\fill}

{\footnotesize 
\centerline{\copyright~\papercopyright. \href{\paperlicenceurl}{\paperlicence}.}}
\vspace*{2mm}

\end{titlepage}

%%%%%%%%%%%%%%%%%%%%%%%%%%%%%%%%
%%%%%  EOD OF TITLE PAGE  %%%%%%
%%%%%%%%%%%%%%%%%%%%%%%%%%%%%%%%

%  empty page follows the title page ----
\newpage
\setcounter{page}{2}
\mbox{~}

%\twocolumn
% %%%%%%%%%%%%% ---------

\renewcommand{\thefootnote}{\arabic{footnote}}
\setcounter{footnote}{0}

\cleardoublepage

%%%%%%%%%%%%%%%%%%%%%%%%%
%%%%% Main text %%%%%%%%%
%%%%%%%%%%%%%%%%%%%%%%%%%

\pagestyle{plain} 
\setcounter{page}{1}
\pagenumbering{arabic}

Over the past decade, analyses of flavour-changing neutral current decays with the quark-level transition \mbox{$\decay{b}{s\mup\mun}$} have shown consistent discrepancies with Standard Model (SM) expectations~\cite{LHCb-PAPER-2014-006,LHCb-PAPER-2016-025,LHCb-PAPER-2016-012,LHCb-PAPER-2020-002,LHCb-PAPER-2021-014,LHCb-PAPER-2023-032,LHCb-PAPER-2023-033,LHCb-PAPER-2024-011,CMS:2024atz,Belle:2016fev}. At the same time, measurements of branching-fraction ratios in tree-level $\decay{b}{c\ell^-{\neub}_{\ell}}$ decays, with final states containing either a tau or a light lepton, $R(D^{(\ast)})$, have shown deviations from lepton flavour universality~\cite{LHCb-PAPER-2024-007,LHCb-PAPER-2017-027,LHCb-PAPER-2017-017,LHCb-PAPER-2015-025,LHCb-PAPER-2022-039,BaBar:2012obs,Belle:2015qfa,Belle:2017ilt,Belle:2019rba,Belle-II:2023aih,Belle-II:2024ami,Belle-II:2025yjp}.  

Models of physics beyond the SM that account for these discrepancies predict enhancements of several orders of magnitude in the branching fractions of flavour-changing neutral current $\decay{b}{s\taup\taum}$ transitions~\cite{Alonso:2015sja,Barbieri:2015yvd,Crivellin:2017zlb,Buttazzo:2017ixm,Capdevila:2017iqn,Bordone:2018nbg,Altmannshofer:2020axr,Aebischer:2022oqe,Crivellin:2019dwb}. Large increases of these branching fractions may also be expected solely within the context of explaining the \mbox{$\decay{b}{s\mup\mun}$} anomalies~\cite{Alguero:2022wkd,Bobeth:2011st}. 

Standard Model predictions have been made for the branching fractions of \mbox{$\decay{\Bz}{\Kstarz\taup\taum}$} and $\decay{\Bs}{\phiz\taup\taum}$ decays that are both of order $10^{-7}$~\cite{Capdevila:2017iqn}.
Throughout this Letter, charge conjugate processes are implied and the symbols \Kstarz and \phiz refer to the specific states $\Kstar(892)^0$ and $\phi(1020)$, respectively.
In a weak effective theory, the coupling strength of the vector (axial vector) operators is given by the Wilson coefficients $C_{9(10)}^{\tau\tau}$. Beyond-the-SM effects may be introduced via a shift, $\Delta$, such that $C_{9}^{\tau\tau} = C_{9}^{\rm SM} - \Delta$ and $C_{10}^{\tau\tau}=C_{10}^{\rm SM}+\Delta$, where $\mathcal{C}^{\rm SM}_{9}$ is $4.1$ and $\mathcal{C}^{\rm SM}_{10}$ is $-4.3$~\cite{Capdevila:2017iqn}. If $\Delta \gg C_{9,10}^{\rm SM}$ the branching fractions are given by~\cite{Capdevila:2017iqn}
\begin{equation}\label{eq:np_brs}
    \begin{split}
        \mathcal{B}(\decay{\Bz}{\Kstarz\taup\taum}) &= (10.1 \pm 0.8)  \times 10^{-9} \times \Delta^2, \\
         \mathcal{B}(\decay{\Bs}{\phiz\taup\taum}) &= (9.1 \pm 0.5) \times 10^{-9} \times \Delta^2. \\
    \end{split}
\end{equation}
Models with $\Delta = O(100)$, corresponding to branching fractions of order $10^{-5}$, are consistent with the current $R(D^{(\ast)})$ results.
The observation of such large branching fractions for these decays would unambiguously indicate the presence of physics beyond the SM. Searches for $\decay{\BdorBs}{\taup\taum}$~\cite{LHCb-PAPER-2017-003}, $\decay{\Bp}{\Kp\taup\taum}$~\cite{BaBar:2016wgb} and $\decay{\Bz}{ \Kstarz\taup\taum}$~\cite{Belle:2021ecr,Belle-II:2025lwo} decays have previously been conducted, resulting in upper limits on the branching fractions of order $10^{-3}$. In the most recent of these searches, the Belle II collaboration constrains \mbox{$\mathcal{B}(\decay{\Bd}{\Kstarz\taup\taum}) < 1.8\times10^{-3}$} at $90\%$ confidence level (CL), reconstructing tau candidates via their decays to single-muon, electron, charged pion and charged $\rho$ final states~\cite{Belle-II:2025lwo}. Searches for rescattering effects in $\decay{\Bd}{\Kstarz\mup\mun}$~\cite{Cornella:2020aoq} constrain $|\mathcal{C}_{9}^{\tau\tau}| < 600$ at $95\%$ CL~\cite{LHCb-PAPER-2024-011}.

In this Letter, two new \lhcb searches for $\decay{b}{s\taup\taum}$ decays are performed using the modes $\decay{\Bz}{\Kp\pim\taup\taum}$ and $\decay{\Bs}{\Kp\Km\taup\taum}$. In both cases the tau leptons are reconstructed using the decay $\decay{\taup}{\mup\neum\neutb}$. 
Although the two searches reconstruct different final states, the analysis strategy is common between them. Reconstructing a final state with two charged hadrons allows the location of the $\BdorBs$ decay vertex to be determined from the intersection of the hadron tracks. 
Knowledge of this vertex location facilitates efficient background suppression and partial reconstruction of the $b$-hadron decay topology. Furthermore, the large branching fraction of the $\decay{\taup}{\mup\neum\neutb}$ decay, and the high efficiency of muon reconstruction at \lhcb~\cite{LHCb-DP-2013-001}, result in large samples of tau candidates for these studies. The searches are conducted using a sample of $pp$ collision data, corresponding to an integrated luminosity of $5.4\invfb$, collected with the \lhcb experiment during 2016--2018.

The dihadron invariant-mass spectra are split into bins to account for regions where the $\Kp\pim$ and $\Kp\Km$ final states may be dominated by resonant or nonresonant contributions (see Table~\ref{tab:Table1}). 
The range of dihadron masses investigated in the two searches spans from the low-mass resonances in each dihadron final state, \Kstarz and \phiz, for $\Kp\pim$ and $\Kp\Km$ respectively, up to their respective kinematic limits, $m_{\BdorBs} - 2m_\tau$, where $m_{\BdorBs}$ is the mass of the \BdorBs meson and $m_{\tau}$ is the mass of the tau lepton~\cite{PDG2024}. 
In order to avoid bias, the results of the analysis were not examined until the full procedure had been finalised. 

\begin{table}[t!]
   \caption{Analysed bins of dihadron mass for the (left) $\Kp\pim$  and (right) $\Kp\Km$ final states.}
\begin{center}
\begin{tabular}{cc}
   \toprule
   \multicolumn{2}{c}{Mass range ($\mevcc$) } \\ 
    \midrule
    $792 < m_{\Kp\pim} < 992$ & $\;\;980 < m_{\Kp\Km} < 1060$ \\
	$\ 992 < m_{\Kp\pim} < 1330$ & $1060 < m_{\Kp\Km} < 1200$ \\
	$1330 < m_{\Kp\pim} < 1530$ &$1200 < m_{\Kp\Km} < 1400$ \\
	$1530 < m_{\Kp\pim} < 1726$ & $1400 < m_{\Kp\Km} < 1600$\\
	-- & $1600 < m_{\Kp\Km} < 1813$ \\
    \bottomrule
  \end{tabular}\end{center}
\label{tab:Table1}
\end{table}

The \lhcb detector~\cite{LHCb-DP-2008-001,LHCb-DP-2014-002} is a single-arm forward spectrometer covering the pseudorapidity range $2 < \eta < 5$, designed for the study of particles containing \bquark\ or \cquark\ quarks. The detector elements that are particularly relevant to this analysis are: a silicon-strip vertex detector surrounding the $pp$ interaction region that allows \bquark\ and \cquark\ hadrons to be identified from their characteristically long flight distance; a tracking system that provides a measurement of the momentum of charged particles; and two ring-imaging Cherenkov detectors that are able to discriminate between different species of charged hadrons. Muons are identified by a
system composed of alternating layers of iron and multiwire proportional chambers. The online event selection, which is the same for both decays studied in this Letter, is performed by a hardware stage and two software stages~\cite{LHCb-DP-2019-001}. The hardware stage requires events to have at least one muon with high transverse momentum. The software stages require a two-, three- or four-track secondary vertex with a significant displacement from all primary $pp$ interaction vertices (PVs).

Simulation is used to model the effects of the detector acceptance and selection requirements on the signal decays and several specific background modes. In the simulation, $pp$ collisions are generated using \pythia~\cite{Sjostrand:2007gs,*Sjostrand:2006za} with a specific \lhcb configuration~\cite{LHCb-PROC-2010-056}. Decays of unstable particles are described by \evtgen~\cite{Lange:2001uf}, in which final-state radiation is generated using \photos~\cite{davidson2015photos}. The interaction of the generated particles with the detector and its response are implemented using the \geant toolkit~\cite{Allison:2006ve, *Agostinelli:2002hh} as described in Ref.~\cite{LHCb-PROC-2011-006}.
The simulated samples are weighted in order to account for known data-simulation differences in the \BdorBs production kinematics and detector occupancy, as well as hardware trigger efficiencies~\cite{LHCb-PUB-2014-039}, track reconstruction efficiency~\cite{LHCb-DP-2013-002} and particle-identification performance~\cite{LHCb-PUB-2016-021}.
Each of these corrections is derived from dedicated control samples selected from the data.

The selection of signal candidates for each of the two final states follows the same strategy, although the precise requirements are not identical. Tracks are required to be of good quality, be displaced from any PV, meet minimum momentum requirements, and have particle-identification criteria consistent with the assigned particle hypotheses. The particle identification is particularly important in order to produce data sets for the two searches with minimal overlap.

Pairs of oppositely charged muons and hadrons are combined to form $b$-hadron candidates.
The dihadron candidates are required to be consistent with originating from a common vertex, significantly separated from the associated PV, and to have a large momentum component transverse to the beam line. The associated PV is that which best aligns with the momentum vector of the (partially reconstructed) $\BdorBs$ candidate.
As the tau particles are expected to fly approximately in the direction of the $b$-hadron, the vertex formed by the intersection of the two muon tracks is required to be located further downstream from the associated PV than the vertex formed by the intersection of the hadron tracks.
In all two- and four-track combinations requirements are placed on the angle between the candidate momentum vector and the line of flight defined by the PV and candidate decay vertex.

Although different specific backgrounds are associated with each final state, the background categories and the strategies to reduce them are common between the two channels. 
A significant source of background consists of decays of the type \mbox{$\decay{\BdorBs}{\Dsm(\to \hadron_1^+\hadron_2^-\mun\neumb)\mup\neum}$}, where $h_{1}^{+}h_{2}^{-}$ represents either $\Kp\pim$ or $\Kp\Km$~\cite{PDG2024}.
A criterion is imposed that the reconstructed  $h_{1}^{+}h_{2}^{-}\mun$ mass must be greater than the known \Dsm mass~\cite{PDG2024}, which effectively rejects this background.
While this reduces the signal efficiency by a factor of approximately two, it reduces to negligible levels an otherwise irreducible set of large background modes.

Another significant background arises from decays of the type $\decay{\Bd}{D^{(\ast)}\Db{}^{(\ast)}\Kstarz}$ or $\decay{\Bs}{D^{(\ast)}\Db{}^{(\ast)}\phiz}$, where both $D$ mesons decay semimuonically. Selection requirements on the missing-mass squared, $m_{\rm miss}^{2}$, and the reconstructed $q^{2}$ reduce this background to a negligible level. The missing-mass squared is defined as \mbox{$m_{\rm miss}^{2} = (P_{\BdorBs} - P_{\hadron_1^+\hadron_2^-} - P_{\mup\mun})^{2}$}, where $P_{\BdorBs}$, $P_{\hadron_1^+\hadron_2^-}$ and $P_{\mup\mun}$ are the reconstructed four-momenta of the $b$-hadron candidate, the dihadron pair and the dimuon pair, respectively.
The reconstructed \qsq is the square of the ditau invariant mass, estimated as $\qsq = (P_{\BdorBs} - P_{\hadron_1^+\hadron_2^-})^{2}$. The $\BdorBs$ four-momentum cannot be determined precisely, as the $\BdorBs$ candidate is partially reconstructed. Therefore an approximation is used that assumes the proper velocity of the reconstructed $\BdorBs$ candidate along the beam line is the true proper velocity~\cite{LHCb-PAPER-2015-025}. 
The resulting estimated quantities have significant discriminating power, despite their limited resolution.  
This is due to the different decay topologies of the signal modes and the backgrounds. 
The particles producing the muons have different masses for signal ($\decay{\taup}{\mup\neum\neutb}$) and background ($\decay{\Dp}{ \mup\neum X}$, where $X$ is any additional particle), resulting in different reconstructed $q^2$ distributions. In addition, the background decays contain fewer unreconstructed neutrinos (two) than signal (four) but can have additional neutral or charged particles such as $\Kz$ mesons, and thus populate different regions of $m_{\rm miss}^{2}$. 
Background modes such as $\decay{\Bz}{K_1^{0}\taup\taum}$, where $K_1^{0}$ refers to $K_1(1270)^0$ and $K_1(1400)^0$ mesons, and $\decay{K_1^{0}}{\Kp\pim X}$ (where $X$ is any particle), have a negligible impact on this analysis given their small branching fractions.

After the selection requirements, the remaining components of the two data sets are in one of four categories: signal, either $\decay{\Bz}{\Kp\pim\taup\taum}$ or $\decay{\Bs}{\Kp\Km\taup\taum}$; combinatorial background, which is formed from random combinations of tracks in the event; semileptonic cascade background, corresponding to decays of the type $\decay{\BdorBs}{\Dzb(\to \hadron_1^+\mun\neumb)\hadron_2^-\mup\neum}$; and a small residual background from final-state particle misidentification. Multiclass boosted decision tree~(BDT)~\cite{Breiman,AdaBoost} classifiers are trained to discriminate between signal, combinatorial background, and semileptonic cascade background. A separate BDT is optimised and trained for each final state and dihadron mass bin. The BDTs are implemented using the \texttt{lightgbm} package~\cite{lightgbm}.  
In training these BDTs, combinatorial background is represented with same-sign muon samples from the data, corresponding to the final state $\hadron_1^+\hadron_2^-\mup\mup$, as this sample is dominated by candidates formed from random combinations of tracks. The semileptonic cascade backgrounds are represented with simulated samples of the relevant decay modes. 
Signal is represented with simulated \mbox{$\decay{\Bd}{ \Kp\pim\taup\taum}$} and \mbox{$\decay{\Bs}{\Kp\Km\taup\taum}$} for the two dihadron modes that are investigated. In each case, the signal is generated uniformly in phase-space. 

A series of features pertaining to the separation of the muons from the dihadron vertex, as well as to the decay topology, are used to separate signal and background. The same features are used for the two searches and in each dihadron mass bin. Of particular importance are the reconstructed tau flight distance and a feature representing how well-isolated the signal candidate is from other tracks in the event~\cite{LHCb-PAPER-2015-013}. The tau flight distance is reconstructed with the assumption that the tau travels approximately in the direction of the $b$ hadron, such that the intersection of the muon line of flight and $b$-hadron line of flight is an estimate of the tau decay vertex. Backgrounds with additional charged particles arise both from partially reconstructed decays of $b$ hadrons, where one or more particles are not reconstructed, and from combinatorial background, in which tracks from several decays are wrongly identified as a candidate. In both cases, efficient reconstruction of the additional charged particles allows these backgrounds to be readily identified and rejected by a specific multivariate algorithm~\cite{LHCb-PAPER-2015-013} which estimates the probability that an additional track in the event stems from the same vertex as the signal candidate.

The analysis sensitivity is affected by the hyperparameters of the BDTs, such as learning rate and maximum depth. Optimal values of these hyperparameters are determined for each final state and dihadron mass bin by maximising a Punzi figure of merit~\cite{Punzi:2003bu} with the \texttt{optuna}~\cite{optuna_2019} python package. The BDTs are trained using k-fold cross-validation with the shapes of the output variables defined only from the test samples~\cite{kFold}. An early stopping procedure is used to avoid overtraining. After the selection procedure, less than $0.1\%$ of remaining events have multiple candidates. In the cases where this happens, one candidate is arbitrarily selected per event. The efficiencies of the BDTs, given the preceding selections, vary between the bins because of the different kinematic distributions of the particles. In the dihadron mass bin including the $\Kstarz$ resonance (corresponding to a $\Kp\pim$ dihadron final state), the signal efficiency is $83\%$, whilst rejecting $95\%$ of combinatorial background and $61\%$ of semileptonic background for the maximal figure of merit. For the $\Kp\Km$ final state in the bin including the $\phiz$ resonance, the maximal figure of merit is found where $87\%$ of signal is kept for $92\%$ combinatorial background and $42\%$ semileptonic background rejection.

There is no suitable mass variable that can be fitted to determine the signal yields, due to the four unreconstructed neutrinos in the final state. An alternative strategy is therefore employed. Candidates are selected that have their BDT classifier score in the signal category higher than in either of the other categories. From these candidates, for each of the two searches, the signal yields in each dihadron mass bin are obtained from independent unbinned extended maximum-likelihood fits to the distribution of the BDT classifier output for the signal category. 
The shapes of each fit component are represented by Gaussian kernel density estimations (KDEs)~\cite{Cranmer:2000du}. Sigmoid functions are used to ensure that the distributions are exactly zero at the upper and lower edges of the BDT-output range. 
The BDT classifier output shapes are derived independently for each fit component and in each dihadron mass bin using either simulation or control samples. The optimal shape for each fit component is found by minimising the likelihood of the relevant proxy sample as a function of the KDE bandwidth. The shapes for the signal modes and semileptonic cascade backgrounds are derived from simulation. The combinatorial shapes are established using the same-sign muon data control samples.
Dedicated data control samples are used to unfold both the shape and the yield of the misidentified background components using the \lhcb PIDCalib package~\cite{LHCb-DP-2018-001} and the \texttt{pyunfold} toolkit~\cite{Bourbeau2018}, as in Ref.~\cite{LHCb-PAPER-2024-007}. The yields of all other fit components are floated. 

\begin{figure}[!tb]
    \centering
    \raisebox{-\height}{\includegraphics[width=0.37\linewidth]{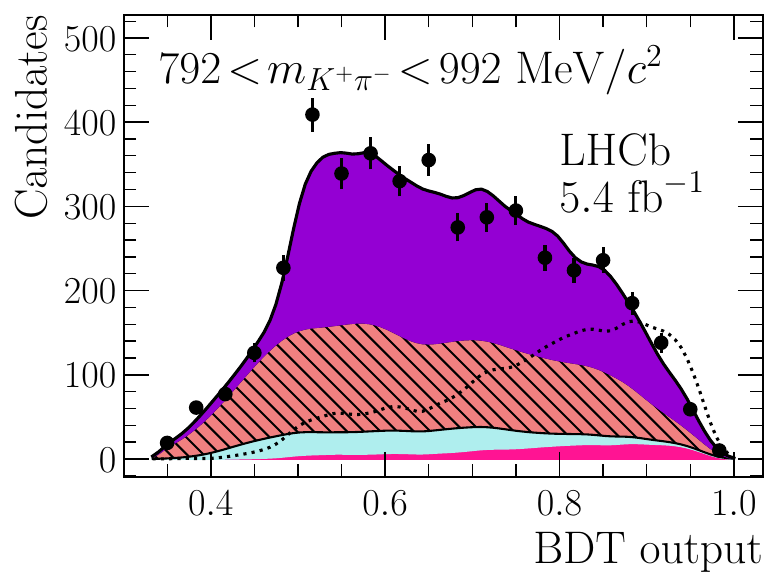}}
    \raisebox{-\height}{\includegraphics[width=0.37\linewidth]{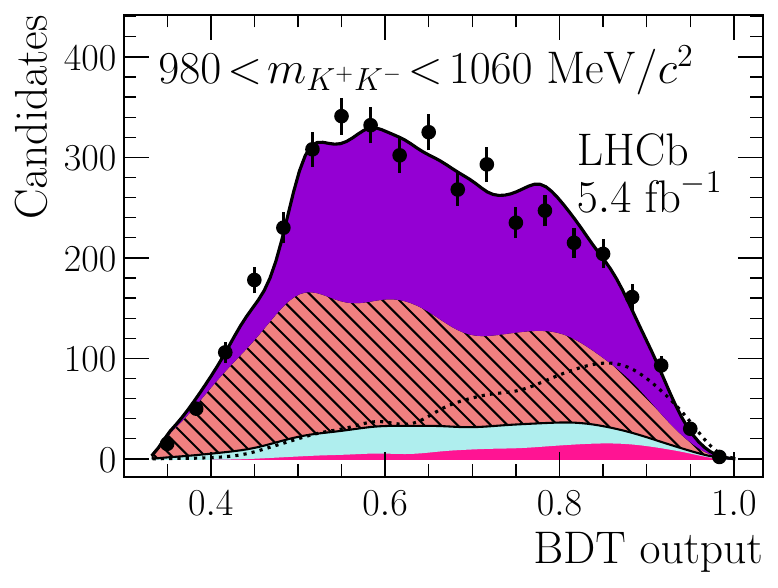}}
    \raisebox{-\height}{\includegraphics[width=0.24\linewidth]{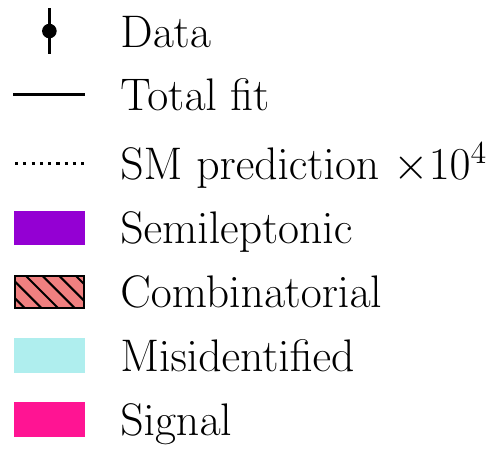}}
    \caption{Distribution of the BDT classifier output for selected candidates of the (left) \mbox{$B^0\to K^+\pi^-\tau^+\tau^-$} and (right) $B^0_s\to K^+K^-\tau^+\tau^-$ data samples in the lowest dihadron mass bin of each. The respective fit results are also shown. The distribution for a hypothetical signal, with branching fraction enhanced by four orders of magnitude compared to the SM prediction, is also shown in each plot for illustrative purposes.}
    \label{fig:Fig1}
\end{figure}

For each of the two final states, the result of the fit to the lowest dihadron mass bin is shown in Fig.~\ref{fig:Fig1}. As the signal yields are not significant in any bin, upper limits are set with the CL$_s$ method~\cite{CLs}. The fit quality for the background-only hypothesis is determined with pseudoexperiments and each bin is found to have a p-value in excess of $5\%$. 

In order to obtain the upper limits on the signal branching fractions, the decays \mbox{$\decay{\Bd}{\jpsi(\to\mup\mun)\Kstarz(\to\Kp\pim)}$} and \mbox{$\decay{\Bs}{\jpsi(\to\mup\mun)\phiz(\to\Kp\Km)}$} are used as separate normalisation modes for the $\decay{\Bz}{\Kp\pim\taup\taum}$ and $\decay{\Bs}{\Kp\Km\taup\taum}$ searches, respectively. 
The branching fraction of one of the signal decays in a given dihadron mass bin $i$, $\mathcal{B}^{\textrm{sig}}_{i}$ is calculated as
\begin{equation}
    \mathcal{B}^{\rm sig}_{i} = \frac{N_{i}^{\rm sig}}{N^{\rm norm}} \times \frac{\epsilon^{\rm norm}}{\epsilon_{i}^{\rm sig}}  \times \mathcal{B}^{\rm norm},
\end{equation}
where $N^{\textrm{sig}}_{i}$ represents the signal yield in bin $i$ and $\epsilon^{\rm sig}_{i}$ is the corresponding signal efficiency. The yield, efficiency and branching fraction for the normalisation mode are denoted by $N^{\rm norm}$, $\epsilon^{\rm norm}$ and $\mathcal{B}^{\rm norm}$ respectively. The efficiencies are derived from simulation. 
The normalisation modes are chosen because their branching fractions are large and well established~\cite{PDG2024} and their reconstructed final states are identical to those of their respective signal modes. 

The initial selection for the two normalisation modes is the same as for the corresponding signal decays, but with the requirements on the reconstructed \qsq and $m_{\rm miss}^{2}$ omitted. 
Separate BDTs are trained for discriminating the normalisation decays from backgrounds, with the same discriminating variables as for the BDTs used to select the signal candidates. However, as the normalisation modes are fully reconstructed, backgrounds from partially reconstructed decays are minimal and so these BDTs  discriminate only the normalisation candidates from combinatorial background. The normalisation yields are obtained from fits to the relevant  $\hadron_1^+\hadron_2^-\mup\mun$ invariant-mass distributions.

The values of the extracted upper limits are impacted by systematic uncertainties. These are accounted for in the determination of the upper limits in the CL$_s$ method by variations of the models used to generate the pseudoexperiments. The most significant systematic uncertainty is due to the limited size of the data control samples used to construct the KDEs in the fits. These effects will be reduced with more data.
The alternative models are made by resampling the proxy samples with replacement~\cite{efron:1979} to construct alternative KDEs.

The choice of KDE hyperparameters also affects the upper limit results. In  order to model the effect of these choices, the pseudoexperiments used for the calculation
of the upper limits with the CL$_s$ method are generated varying the bandwidths within two standard deviations of their optimal value, both with and without sigmoid functions.

A further source of uncertainty arises from the phase-space model used to simulate the two signal decay modes. 
Weights are applied to emulate alternative physics models, using predictions from Refs.~\cite{Seidel:2004jh,Beneke:2003az,Descotes-Genon:2015hea,Inami:1980fz,Beneke:2004dp,Asatrian:2003vq,Straub:2015ica,Bobeth:2013uxa} implemented with the \texttt{flavio}~\cite{straub:2018flavio} package. 
The changes in BDT output shapes and in the overall signal efficiencies are minimal when physics models with $\mathcal{O}(10^2)$ shifts in the Wilson coefficients $\mathcal{C}^{\tau\tau}_9$ and $\mathcal{C}^{\tau\tau}_{10}$ are taken into account, and when incorporating uncertainties due to the simulation corrections.

The analysis method is validated by analysing several decay modes with the same final states as the signals under study.
The branching fractions of $\decay{\Bs}{\Kp\Km\mup\mun}$ and $\decay{\Bd}{\Kp\pim\mup\mun}$ are obtained with the fits of an appropriate BDT classifier score and compared to previous measurements~\cite{LHCb-PAPER-2021-014, LHCb-PAPER-2016-012}. The results extracted with this method are found to be consistent with the published numbers.

\begin{figure}[!tb]
    \centering
    \includegraphics[width=0.49\linewidth]{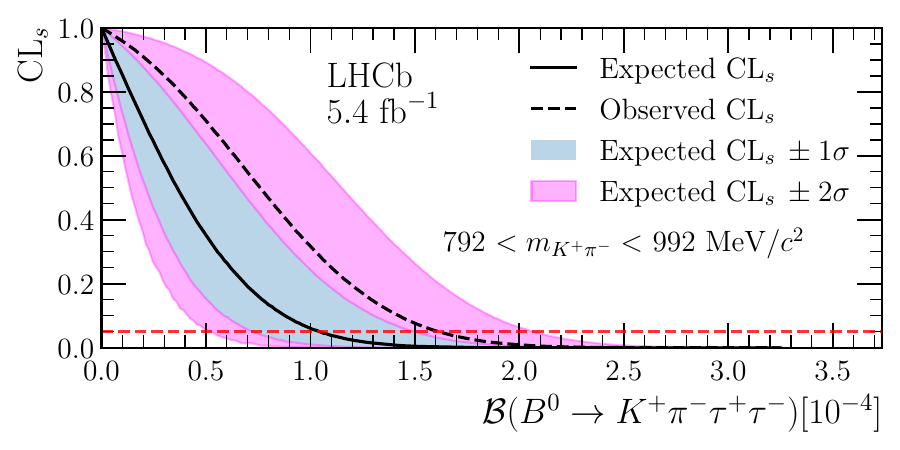}
    \includegraphics[width=0.49\linewidth]{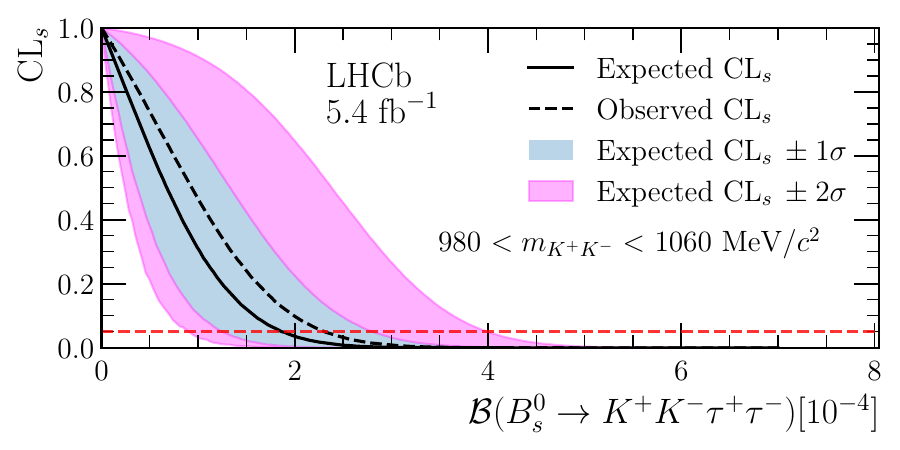}
    \caption{Expected and observed upper limits obtained with the CL$_s$ method for the lightest bins containing (left) $K^+\pi^-$ and (right) $K^+K^-$ final states.}
    \label{fig:Fig2}
\end{figure}

The resulting CL$_s$ upper limits for $\decay{\Bd}{\Kp\pim\taup\taum}$ and $\decay{\Bs}{\Kp\Km\taup\taum}$ in each of the dihadron mass bins are given in Table~\ref{tab:Table2}, and the CL$_{s}$ scans for the lightest dihadron mass bins are shown in  Fig.~\ref{fig:Fig2}. The $\decay{\Bd}{\Kp\pim\taup\taum}$ results are recast as a search for the decay $\decay{\Bs}{\Km\pip\taup\taum}$,  which proceeds via a \mbox{$\decay{b}{d\taup\taum}$} transition, using the ratio of hadronisation fractions $\frac{f_{s}}{f_{d}}$~\cite{LHCb-PAPER-2020-046}. The resulting upper limits are also given in Table~\ref{tab:Table2} and are the first for this process.

\begin{table}[tb]
  \caption{Upper limits on the branching fractions of $\decay{\Bd}{\Kp\pim\taup\taum}$, $\decay{\Bs}{\Kp\Km\taup\taum}$ and $\decay{\Bs}{\Km\pip\taup\taum}$ decays in the analysed bins of dihadron mass at $90\%$ and $95\%$ CL. 
  }
\begin{center}\begin{tabular}{c c c c c c}
    \toprule
    Confidence level  & \multicolumn{4}{c}{Upper limit on $\mathcal{B}(\decay{\Bd}{\Kp\pim\taup\taum})$} & \\
    \midrule 
    $m_{\Kp\pim}$ ($\mevcc$) & $[792, 992]$ & $[992, 1330]$ & $[1330, 1530]$ & $[1530, 1726]$ & \\
    \midrule 
   $90\%$ & $1.4\times10^{-4}$ & $2.7\times10^{-5}$ & $1.0\times10^{-5}$ & $ 2.7\times10^{-6}$  &  \\
   $95\%$ & $1.6\times10^{-4}$ & $3.4\times10^{-5}$ & $1.1\times10^{-5}$ & $ 3.3\times10^{-6}$ & \\
    \midrule
    \midrule
    Confidence level  & \multicolumn{4}{c}{Upper limit on $\mathcal{B}(\decay{\Bs}{\Kp\Km\taup\taum})$} &  \\
    \midrule 
   $m_{\Kp\Km}$ ($\mevcc$) &  $[980, 1060]$ & $[1060, 1200]$ & $[1200, 1400]$ & $[1400, 1600]$ & $[1600, 1813]$  \\
     \midrule 
   $90\%$ & $2.0\times10^{-4}$ & $1.3\times10^{-4}$ & $1.2\times10^{-4}$ & $6.8\times10^{-5}$ & $3.2\times10^{-5}$ \\
   $95\%$ & $2.3\times10^{-4}$ & $1.5\times10^{-4}$ & $1.4\times10^{-4}$ & $7.6\times10^{-5}$ & $3.6\times10^{-5}$ \\
    \midrule
    \midrule
    Confidence level  & \multicolumn{4}{c}{Upper limit on $\mathcal{B}(\decay{\Bs}{\Km\pip\taup\taum})$} & \\
    \midrule 
    $m_{\Kp\pim}$ ($\mevcc$) & $[792, 992]$ & $[992, 1330]$ & $[1330, 1530]$ & $[1530, 1726]$ & \\
    \midrule 
   $90\%$ & $6.5\times10^{-4}$ & $1.2\times10^{-4}$ & $5.1\times10^{-5}$ & $1.7\times10^{-5}$ &  \\
   $95\%$ & $7.3\times10^{-4}$ & $1.5\times10^{-4}$ & $6.2\times10^{-5}$ & $2.1\times10^{-5}$ &  \\
    \bottomrule
  \end{tabular}\end{center}
\label{tab:Table2}
\end{table}

The upper limit for the $\decay{\Bz}{\Kp\pim\taup\taum}$ decay in the $792< m_{\Kp\pim}<992\mevcc$ bin is reinterpreted as a limit on the $\decay{\Bz}{\Kstarz\taup\taum}$ decay by assuming that the \Kstarz resonance is the only contribution, and accounting for the \mbox{$\decay{\Kstarz}{\Kp\pim}$} branching fraction~\cite{PDG2024} and the fraction of the \Kstarz resonance that lies within this mass range. Similarly, the $\decay{\Bs}{\Kp\Km\taup\taum}$ search in the \mbox{$980< m_{\Kp\Km}<1060\mevcc$} bin is reinterpreted as a search for the decay $\decay{\Bs}{\phi\taup\taum}$. The results of these reinterpretations are
\begin{equation*}
    \begin{split}
    \mathcal{B}(\decay{\Bd}{\Kstarz\taup\taum}) &< 2.8 \times10^{-4} ~(2.5\times10^{-4}) \; \mathrm{ at } \; 95\% ~(90\%) \; \mathrm{ CL},\\
    \mathcal{B}(\decay{\Bs}{\phiz\taup\taum}) &< 4.7\times10^{-4} ~(4.1\times10^{-4})  \; \mathrm{ at } \;  95\% ~(90\%) \;\mathrm{ CL}.\\
    \end{split}
\end{equation*}
Combining these upper limits and the model and assumptions from Ref.~\cite{Capdevila:2017iqn}, limits are also set on the square of the shift $\Delta$ in the Wilson coefficients $\mathcal{C}^{\tau\tau}_9$ and $\mathcal{C}^{\tau\tau}_{10}$, defined in Eq.~\ref{eq:np_brs}. The resulting upper limits are shown in Table~\ref{tab:Table3}. 
These are tighter constraints than those obtained in Ref.~\cite{LHCb-PAPER-2024-011} by more than a factor of three.

\begin{table}[!tb]
  \caption{
  Upper limit on $\Delta^2$ at $90\%$ and $95\%$ CL.}
\begin{center}\begin{tabular}{c c c  }
    \toprule
    Confidence level & $\decay{\Bd}{\Kp\pim\taup\taum}$ & $\decay{\Bs}{\Kp\Km\taup\taum}$ \\
    \midrule 
   $90\%$ & $2.5\times10^4$ & $4.5\times10^4$  \\
   $95\%$ & $2.9\times10^4$ & $5.2\times10^4$ \\

    \bottomrule
  \end{tabular}\end{center}
\label{tab:Table3}
\end{table}

In summary, this Letter reports searches for the $\decay{b}{ s\taup\taum}$ decays $\decay{\Bz}{\Kp\pim\taup\taum}$ and $\decay{\Bs}{\Kp\Km\taup\taum}$. 
No significant signal is observed for either decay and upper limits on the respective branching fractions are set. Interpreting the former as a search for the $\decay{\Bz}{\Kstarz\taup\taum}$ decay leads to an upper limit on the branching fraction of this decay that is an order of magnitude better than the current world's best~\cite{Belle-II:2025lwo}. The searches for the \Bs decay mode are the world's first. 
Given the anomalies observed in the measurements of $R(D^{(\ast)})$, a range of new physics models have been constructed that predict significant enhancements in the branching fractions of these decay modes.
This analysis augurs a comprehensive programme of $\decay{b}{ s\taup\taum}$ searches at \lhcb that will challenge such new-physics models with the data collected in the ongoing Large Hadron Collider run.

\hfill \break

\section*{Acknowledgements}
%
% These Acknowledgements valid from 12/09/25
%
\noindent We express our gratitude to our colleagues in the CERN
accelerator departments for the excellent performance of the LHC. We
thank the technical and administrative staff at the LHCb
institutes.
We acknowledge support from CERN and from the national agencies:
ARC (Australia);
CAPES, CNPq, FAPERJ and FINEP (Brazil); 
MOST and NSFC (China); 
CNRS/IN2P3 (France); 
BMFTR, DFG and MPG (Germany);
INFN (Italy); 
NWO (Netherlands); 
MNiSW and NCN (Poland); 
MCID/IFA (Romania); 
%MSHE (Russia); 
MICIU and AEI (Spain);
SNSF and SER (Switzerland); 
NASU (Ukraine); 
STFC (United Kingdom); 
DOE NP and NSF (USA).
%%%%%%%%%%%%%%%%%%%%%%%%%%%%%%%%%%%%%%%%%%%%%
We acknowledge the computing resources that are provided by ARDC (Australia), 
CBPF (Brazil),
CERN, 
IHEP and LZU (China),
IN2P3 (France), 
KIT and DESY (Germany), 
INFN (Italy), 
SURF (Netherlands),
Polish WLCG (Poland),
IFIN-HH (Romania), 
%RRCKI and Yandex LLC (Russia), 
PIC (Spain), CSCS (Switzerland), 
and GridPP (United Kingdom).
%%%%%%%%%%%%%%%%%%%%%%%%%%%%%%%%%%%%%%%%%%
We are indebted to the communities behind the multiple open-source
software packages on which we depend.
%%%%%%%%%%%%%%%%%%%%%%%%%%%%%%%%%%%%%%%%%%
Individual groups or members have received support from
%ARC and ARDC (Australia); % moved to national 16/01
Key Research Program of Frontier Sciences of CAS, CAS PIFI, CAS CCEPP, 
Fundamental Research Funds for the Central Universities,  and Sci.\ \& Tech.\ Program of Guangzhou (China);
Minciencias (Colombia);
EPLANET, Marie Sk\l{}odowska-Curie Actions, ERC and NextGenerationEU (European Union);
A*MIDEX, ANR, IPhU and Labex P2IO, and R\'{e}gion Auvergne-Rh\^{o}ne-Alpes (France);
%RFBR, RSF and Yandex LLC (Russia);
Alexander-von-Humboldt Foundation (Germany);
ICSC (Italy); 
%GVA, XuntaGal, GENCAT, Inditex, InTalent and Prog.~Atracci\'on Talento, CM (Spain);
Severo Ochoa and Mar\'ia de Maeztu Units of Excellence, GVA, XuntaGal, GENCAT, InTalent-Inditex and Prog.~Atracci\'on Talento CM (Spain);
SRC (Sweden);
the Leverhulme Trust, the Royal Society and UKRI (United Kingdom).

\clearpage

%\input{justification}

%\clearpage
%\input{supplementary-app}

\addcontentsline{toc}{section}{References}
%\setboolean{inbibliography}{true}
\bibliographystyle{LHCb/LHCb}
\bibliography{main,LHCb/standard,LHCb/LHCb-PAPER,LHCb/LHCb-CONF,LHCb/LHCb-DP,LHCb/LHCb-TDR}

\newpage
% LHCb collaboration author list
% Data extracted on April 13th, 2026 at 5:42pm for paper reference LHCb-PAPER-2025-048
\centerline
{\large\bf LHCb collaboration}
\begin
{flushleft}
\small
R.~Aaij$^{38}$\lhcborcid{0000-0003-0533-1952},
A.S.W.~Abdelmotteleb$^{58}$\lhcborcid{0000-0001-7905-0542},
C.~Abellan~Beteta$^{52}$\lhcborcid{0009-0009-0869-6798},
F.~Abudin\'en$^{58}$\lhcborcid{0000-0002-6737-3528},
T.~Ackernley$^{62}$\lhcborcid{0000-0002-5951-3498},
A. A. ~Adefisoye$^{70}$\lhcborcid{0000-0003-2448-1550},
B.~Adeva$^{48}$\lhcborcid{0000-0001-9756-3712},
M.~Adinolfi$^{56}$\lhcborcid{0000-0002-1326-1264},
P.~Adlarson$^{86}$\lhcborcid{0000-0001-6280-3851},
C.~Agapopoulou$^{14}$\lhcborcid{0000-0002-2368-0147}
% ...

{\footnotesize \it

% ...
$^{14}$Universit{\'e} Paris-Saclay, CNRS/IN2P3, IJCLab, Orsay, France\\
% ...
$^{38}$Nikhef National Institute for Subatomic Physics, Amsterdam, Netherlands\\
% ...
$^{48}$Instituto Galego de F{\'\i}sica de Altas Enerx{\'\i}as (IGFAE), Universidade de Santiago de Compostela, Santiago de Compostela, Spain\\
% ...
$^{52}$Physik-Institut, Universit{\"a}t Z{\"u}rich, Z{\"u}rich, Switzerland\\
% ...
$^{56}$H.H. Wills Physics Laboratory, University of Bristol, Bristol, United Kingdom\\
% ...
$^{58}$Department of Physics, University of Warwick, Coventry, United Kingdom\\
% ...
$^{61}$School of Physics and Astronomy, University of Glasgow, Glasgow, United Kingdom\\
$^{62}$Oliver Lodge Laboratory, University of Liverpool, Liverpool, United Kingdom\\
% ...
$^{70}$Syracuse University, Syracuse, NY, United States\\
% ...
$^{86}$Department of Physics and Astronomy, Uppsala University, Uppsala, Sweden, associated to $^{61}$\\
% ...
}
\end{flushleft}

\end{document}